\documentstyle[%
twoside]{article}

\newcommand{\cabeza}{ Dispersion }
\makeatletter
\renewcommand{\@oddhead}
             { \cabeza \hfill \thepage}
\renewcommand{\@evenhead}
             {\thepage \hfill \cabeza \hfill S. A. Choro\v{s}avin }
\renewcommand{\@oddfoot}{}
\renewcommand{\@evenfoot}{}
\makeatother
\newenvironment{Thm}[2]{\par\addvspace{\bigskipamount}{\bf #1#2}\it }%
{\par\addvspace{\bigskipamount} }

{\par\hspace*{\fill}$\Box$ \par\addvspace{\smallskipamount} }

\author{S.A.~Choro\v{s}avin}
\title{ 
 What do quantum particles do, being under potential barrier?
 Tunnelling time.
 A Virtual Experiment Standpoint.
 }
\date{}
\begin{document}
\maketitle 
\begin{abstract}

\addvspace{1\bigskipamount}\par\noindent
 Addressed, mainly: postgraduates and related readers.

\addvspace{1\bigskipamount}\par\noindent
 Subject:
 Given two classical mechanical 
$1D$-moving particles (material points),
 with identical initial data, 
 one of those particles given free
 and another given to pass through a symmetrical force-barrier,
 a retardation effect is observed:
 After the barrier has been passed over,
 the second particle moves with the same 
 velocity as the free particle, but spacially is retarded 
 with respect to the latter.  
 If the "non-free" particle moves through a potential well, 
 then the retarded particle is the free particle.

\addvspace{1\bigskipamount}\par\noindent
 The question is.
 What phenomena of a similar kind could one find,
 if the $1D$-moving particles were quantum ones?  
 And just what do quantum particles do, being under potential barrier?
 I here say "quantum" in a mathematical sense:  "Schroedinger".

\addvspace{1\bigskipamount}\par\noindent
 To answer the question, I had constructed some suitable 
 Virtual Devices (Java applets) 
 and then for some time experimented various situations.
 I detected some phenomena that I could name "a retardation", however 
 I doubt whether that term is proper.
 Some of those Java applets are available at   
 \verb"http://choroszavin.narod.ru/vlab/index-2.htm"

\addvspace{1\bigskipamount}\par\noindent
\end{abstract}

\newpage

\addvspace{5\bigskipamount}\par\noindent
 Given two classical mechanical 
$1D$-moving particles (material points),
 with identical initial data, 
 one of those particles given free
 and another given to pass through a symmetrical force-barrier:
$$
   F = \left\{ 
   \begin{array}{rcl}
    -F_0 &,& x\in[-w/2+x_0, x_0]
\\
     F_0 &,& x\in[x_0, x_0+w/2]
   \end{array}\right.
\qquad ,
$$
 a retardation effect is observed:
 After the barrier has been passed over,
 the second particle moves with the same 
 velocity as the free particle, but spacially is retarded 
 with respect to the latter.  
 If the "non-free" particle moves through a potential well, 
 then the retarded particle is the free particle.

\addvspace{2\bigskipamount}\par\noindent
 The question is.
 What phenomena of a similar kind could one find,
 if the $1D$-moving particles were quantum ones?  
 And just what do quantum particles do, being under potential barrier?
 I here say "quantum" in a mathematical sense:  "Schroedinger".

\addvspace{2\bigskipamount}\par\noindent
 To answer the question, I had constructed some suitable 
 Virtual Devices (Java applets) 
 and then for some time experimented various situations.
 I detected some phenomena that I could name "a retardation", however 
 I doubt whether that term is proper.
 As for the most certain and definite manifestations of the retardation,  
 I had untill now detected them by those of my Virtual Devices,
 which I had constructed on principles of the stationary scattering theory. 

\addvspace{1\bigskipamount}\par\noindent
 Mathematically they are based on an analysis of 
\begin{eqnarray*}
   i\partial_t \psi(t,x)
&=& 
   -\partial_x^2 \psi(t,x)
              +\sum_a \alpha_{a}\delta(x-x_a) \psi(t,x)(x_a) \,.
\end{eqnarray*}
 A solution (an eigensolution) is of the form 
\begin{eqnarray*}
   \psi(t,x)
&=& 
    e^{-i\omega t} \widetilde{w}_{k}(x) \,,
\quad \omega = k^2 \,,
\end{eqnarray*}
\begin{eqnarray*}
   \widetilde{w}_{k}(x)
&=& 
   e^{ik x} + \sum_a coef_a \cdot e^{i |k| \cdot |x-x_a|} \,,
\end{eqnarray*}
 and any finite linear combination 
\begin{eqnarray*}
   \psi(t,x)
&=& 
   \sum_m A_m e^{i\phi_m} e^{-i\omega_m t} \widetilde{w}_{k_m}(x) \,,
\end{eqnarray*}
 is a solution as well.
 Of course, 
$\{coef_a\}_a$
 depend on 
$\{x_a\}_a$, $\{\alpha_a\}_a$, and $k$ (resp., $k_m$).
 The stationary scattering theory says that if the "free" system 
 behaves as 
\begin{eqnarray*}
   \psi_{free}(t,x)
&=& 
   \sum_m A_m e^{i\phi_m} e^{-i\omega_m t} e^{ik_m x} \,,
\end{eqnarray*}
 then the "non-free" will behave as 
\begin{eqnarray*}
   \psi_{non-free}(t,x)
&=& 
   \sum_m A_m e^{i\phi_m} e^{-i\omega_m t} \widetilde{w}_{k_m}(x) \,.
\end{eqnarray*}
 I took, mainly, 
\begin{eqnarray*}
   \psi_{free}(t,x)
&=& 
   \sum_m e^{-i\omega_m t} e^{ik_m x} \,,
\end{eqnarray*}
 and observed the graphs of 
$$
 |\psi_{free}(t,x)|^2  \,,\quad |\psi_{non-free}(t,x)|^2  \,,
$$
 especially, ones of 
$$
 |\psi_{free}(0,x)|^2  \,,\quad |\psi_{non-free}(0,x)|^2  \,.
$$
 In such a case 
$$
 |\psi_{free}(0,x)|^2 
$$
 looks like a series of quasiperiodic relatively localized excitations. 
 The structure of 
$$
 |\psi_{non-free}(0,x)|^2 
$$
 looks much more complicated, although sometime it could be specified as 
 a set of "excitations" 
 and there I can not formulate my impressions more defintely.
 
\addvspace{1\bigskipamount}\par
 As for the basic theme of my paper, I found that:

 the less was the effective width of the "free"-excitations with respect to 
 the region of 
$$
  \sum_a \alpha_{a}\delta(x-x_a) \psi(t,x)(x_a) \,,
$$
 ---i.e., with respect to 
$$
 \max\{x_a\}_a  - \min\{x_a\}_a
$$
 ,---
 and the less were the distances between scattering centres with respect to
 the effective width of the "free"-exitations,
 the more explicitly I observed a phenomenon which I could name "retardation".
 
\addvspace{1\bigskipamount}\par\noindent
 Some of my Virtual Laboratories (Java applets)
 where I observed described phenomena certainly and explicitly are located at  
 \verb"http://choroszavin.narod.ru/vlab/index-2.htm"

\addvspace{1\bigskipamount}\par\noindent
 An instruction: to observe a retardation phenomemon move the scrollbar named 
$cc\_b$
\par\noindent
 As for "I doubt whether that term is proper", 
 another instruction is: set \verb"n_de_las_ondas=1" and repeat measurements.

\newpage

\addvspace{2\bigskipamount}\par\noindent
{\bf Notes: }

\addvspace{1\bigskipamount}\par\noindent
 The basic theme of the paper is generated
 by a theme of the papers of N.L. Chuprikov.
 The philosophy of the computations has associated 
 with the paper [10] of S.R. Foguel 
 and my papers [6]-[9].

\addvspace{2\bigskipamount}\par\noindent
{\bf Acknowledgements: }

\addvspace{1\bigskipamount}\par\noindent
 No financial support.

\addvspace{1\bigskipamount}\par\noindent
 Many thanks to Wolfgang Christian, Peter Krahmer, David J. Eck and 
 Angel Franco Garcia 
 for the classes. 

\addvspace{1\bigskipamount}\par\noindent
\bibliographystyle{unsrt}

\end{document}